\documentstyle[floats,twocolumn,pre,aps]{revtex}
\input boxedeps
\input epsf
\begin{document}
\title{Spontaneous curvature-induced pearling instability}
\author{Sahraoui Cha\"{\i}eb \thanks{Present address: Departamento de F\'{\i}sica de la
Universidad de Santiago, Av. Ecuador 3493, Casilla 307 Correo 2, Santiago, Chile} and Sergio Rica}
\address{
 Laboratoire de Physique Statistique, Ecole Normale Sup\'erieure
24 rue Lhomond, 75231 Paris Cedex 05, France}

\vskip .2cm
\author{\parbox{430pt}{\vglue 0.3 cm \small
{We investigate the instability of a tubular fluid membranes made of a water
soluble surfactant. The tubules are obtained at high brine salinities. The instability is
due to introduction
within the vesicle multilayer an alkane (Dodecane).
We measure the wavelength of this instability versus
the unperturbed radius of the tubules. To interpret this dependance
we use a model that include not only the surface tension in
the elastic energy, but the spontaneous curvature as well.
The spontaneous curvature is induced by the presence of the oil
in the bilayer of the membranes, and give a selection of a non-zero
wavelength.}
}}
\author{\parbox{430pt}{\vglue 0.3 cm \small PACS numbers:
68.10.-m, 02.40.-k, 47.20.-k, 87.22.Bt,
}}
\maketitle
\protect
\section{Introduction}
Vesicles naturally exist in different shapes. The most common
shapes are spherical and oblates like vesicles, toroidal vesicles
also exist under some particular conditions
\cite{lipow,xavier}.
Tubular vesicles are also common in nature, but they are not
an equilibrium solution for the elastic energy \cite{helfrich}.
One of the most fascinating phenomena are the shape changes of
cells and the manifold of local instabilities of the cell plasma
membrane in particular and the vesicles membranes in general.
Under physiological condition the cell (vesicle) exhibits the
familiar biconcave shape (the discoide form). Removal of some
particular pro\-teins leads to the echino\-cyte form, or to
the stomato\-cyte form \cite{sack1}. The interest of vesicles as  model system of
living cells, for drug delivery or for different industrial usage...etc.
is the origin of several fundamental studies over a couple of decades \cite{madox}.
Local instabilities play an essential role for ma\-terial trans\-port process between
ce\-llular com\-partments or through the pla\-sma mem\-brane.
A most important example is the transport of newly synthesized membranes
to the cell envelope, and it involves three steps: a)~budding of vesicles,
b)~their fission from a parent membrane, c)~their fusion with the target
com\-partment.
Tubular vesicles when excited by mean of different techniques
(UV radiation, optical tweezers...), exhibit curious behavior similar to the
Rayleigh instability in liquid column. This behavior is known as ``pearling''
\cite{moses,rings,sack}. After an exposition to radiation, tubules become
a succession of spheres, separated by narrow necks.
In the case of UV radiation, a long time radiation induces a separation of the spheres
\cite{rings}. When the tubules are excited by optical tweezers,
the shape of the spheres depends on the intensity of the
laser and a long time exposition of the optical tweezers on the
tubules induces a separation of the spheres separated by
narrow necks \cite{moses}. Also it was found that vesicles can suffer
the same pearling instability when subjected to a change of the water Ph
between the inside and the outside of the vesicle \cite{farge}.

In this article we study an instability that suffer tubular vesicles made of a water
soluble amphiphile. This instability is similar to the ones cited above. However, the tubules
are not excited via an irradiation, but the interlayer distance may be affected by
incorporating a solvent of the aliphatic chaines of the amphiphiles into the layer.
In Section II we will describe the experiment and in 
section III we present the theoretical model of the instability observed and finally a
conclusion is presented.
\protect
\section{Experiment}
It is known that Alkanes are more or less good solvents, depending on their chain length,
for the aliphatic chain of the surfactant called AOT (sodium di-2-ethylhexyl-sulfosuccinate)
at the oil-water interface\cite{hamid}. 
AOT is a water-soluble
surfactant, it forms micelles when dissolved in water up to a
concentration equal to 30 mmol/L. Beyond this ``solubility limit'',
AOT forms spherical and multilamellar vesicles \cite{mathese}. These vesicles are 
found to be opaque when observed through a microscope and present different layer
when observed under a phase contrast micropscope. We believe
they are onions like vesciles. However at low salinities and at
low AOT concentrations, it forms unilamelar sphericals
and prolate vesicles as well as dumbell shape vesicles.
At high salinities (we investigated the range 0.1-0.175 mol/L),
we found that AOT in brine forms tubular vesicles \cite{mathese}. These
tubular vesicles can be very long and can be of several order of magnitudes of
their diameter. There are two ways of preparing this phase: Either solubilising AOT
in a very short alkane (Heptane), and drying up the solution before
adding water on the residue left after the evaporation of Heptane;
or mixing first AOT and salt in a ration 1:3 in weight, then pouring
gently the appropriate quantity of water to the AOT/Salt mixture to
have the desired (AOT/Salt) concentrations. For the purpose of this
work we used the second procedure, so we can avoid any kind of
mixing of the two alkanes. If we prepare a low salinity solution by dissolving
the amount of AOT in water and adding brine to bring the solution to the desired salinity, 
no vesicle is observed under the phase contrast microscope. The formation of these vesicles
is probably due to an electrostatic effect between the amphiphiles heads \cite{free}.
In this article we will not describe the
phase observed when changing salinities or amphiphiles concentration.
Instead we will focus on a particular effect of dodecane on these tubules obtained
at high salinities. An instability is studied and the wavelength selected at the onset
of the instability is measured as a function of the radius of the unperturbed tubules.
The selection is found to be a result of the incorporation of the oil into the layer of
the tubule which causes the spontaneous curvature to deviate from zero.

\begin{figure}[tbh]
\centerline{ \epsfxsize=\columnwidth \epsfysize=60 mm
\epsfbox[36 228 576 543]{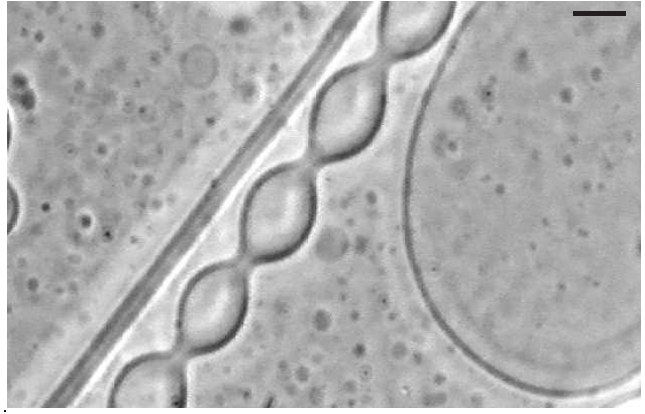}}
\caption
{A peristaltic state, in a thick cylinder. The bar represents 10 $\mu m$.}
\label{epais}
\end{figure}

The vesicles are observed through an inverted phase
contrast microscope ({$40\times$} objective, Nikon diaphot 200).
The cell where the vesicles are observed is a 1-mm-~thick
glass cell, the gap between these two glass slides is also
1-mm-~thick. The AOT-brine solution is left for a couple of days,
so that any shape due to macroscopique flow -after transfering the
solution from a test tube to the observation cell- desapears,
though no shape transformation was observed after the transfer of the
vesicle dispersion into the observation cell.
The brine salinities used corresponding to the minimun of oil-water
interfacial tensions and are 0.175 mol/L for dodecane and 0.075 mol/L for
decane and an AOT concentration of 7.5 mmol/L and 4 mmol/L respectively.
The CMC of the same system without salt is around 2.5 mmol/L.
At 0.075 mol/L of NaCl, the stable shapes are essentially  prolate and spheres;
and at 0.175 mol/L of NaCl, the shapes are cylinders. This shape
transformation from low salinities to high salinities is probably du to an
electrostatic effect, leading to shape transformation)
\cite{fourcade,seifert}.
We observed the effect of dodecane on stable tubules at 0.175 mol/L
(at this salinity the Dodecane-Brine interface tension is minimum) \cite{hamid}.
The introduction of the oil into the cell does not perturbe
hydrodynamically the solution and it is a noninvasive way to
introduce the drop into the solution. To avoids instabilities such as Marangoni
effect, the oil is introduced through a less than $1
\mu m$ diameter orifice. The drop,  with a volume of the order of $0.25 \mu L$, 
travells through the glasse by capillarity before it reaches the solution. The orifice 
is situated at the lateral wall of the cell, that is between the lower glass slide and the 
piece separating the two slides. The drop is then directly introduced into the solution.
Due to the weak solubility of dodecane in water, and its slow diffusion, it takes
several hours for the instability to start. Most probably  the oil travells
inside the solution after being incorporated in the swollen micelles. After the
oil reaches the tubules situated far enough from the point where the drop was
introduced, tubules become unstable, by forming pearls similar to the ones
observed in Ref. \cite{moses,farge}. We have not noticed any shape changes when
the gap  between the horizontal walls of the observation cell is much less than 1
$mm$. Up to now, we are not able to know why this effect depends on the gap
between the cells-walls where the vesicles are being observed. A sinusoidal
instability is initiated, and develops to a peristaltic state, with a reduction
in fluctuations. However large cylinders are stables, this will be clarified in the 
following model. The  structure observed is
periodic and typical necks between the pearls are apparent.  In the case of
thick walled vesicles, the pearls don't disconnect. However tubules with thin
layers, cut into separated spheres at the end of the instability. The time over
which  the spheres diconnect is still unknown. This kind of state is shown in
(fig.~\ref{epais}).

Bar-Ziv and Moses \cite{moses} suggested, after an experiment
where a tubular phospholipidic membrane was excited using an optical tweezers,
that, the instability appears because of a change in the surface tension induced
by the tweezers, by analogy to the Rayleigh instability of a column of liquid
\cite{chandra,granek,gold}.
The surface energy should be the source of
instability, however as the surface energy increases the linear
analysis show that the instability develops only longwave modulations
without any wavelength selection. We will show in the following that in the case
of our experiment surface tension is not sufficient to select a well-defined
periodic spatial modulation of tubular membranes. Instead, in our case an extra length
scale becomes important in the problem. We introduce a simple model which introduce
the physical meaning of this length scale.

\begin{figure}[tbh]
\centerline{\epsfxsize =\columnwidth \epsfysize=\columnwidth\epsfbox{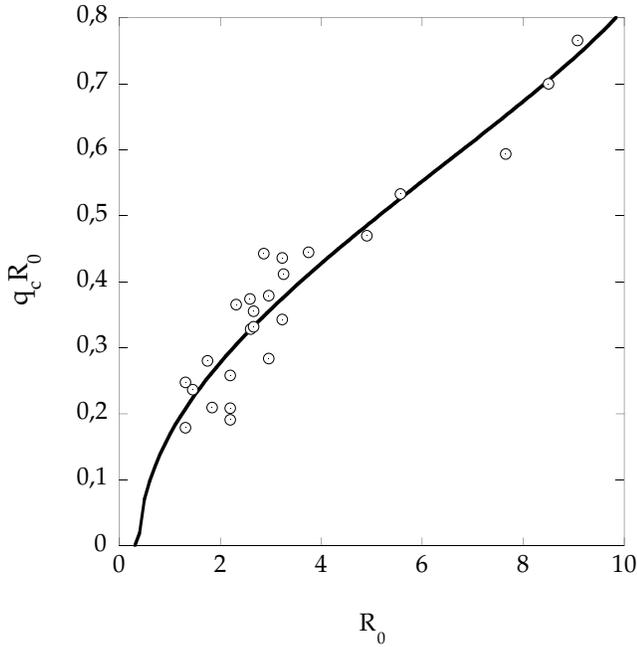}}
\caption
{The product ($q_cR_0$) as a function of the unperturbed radius ($R_0$, in
microns). The line represents a fit with the theory, see later. }
\label{adda1}
\end{figure}

Figure (\ref{adda1}), shows the dependance of the dimensionless
number ($q_cR_0$) as a function of the initial radius of the tubule $R_0$. The
wavelength length at the onset hence $q_c$, is measured whenever a tubule
starts to suffers the instability. The dispersion in the data in
figure~\ref{adda1} is probably due to the fact that the wavelength is not
measured exactly at the onset with an error of the order of 2 seconds (that is after
the establishement of the perstaltic state). The oil molecules reaches
the tubules in a random way, so the onset is also random. The finer
tubules start to suffer the instability earlier than the larger ones, this fact
explains also the dispersion in the data of figure~\ref{adda1}.

The characteristic slope of the experimental values of figure~\ref{adda1},
represents an extra length (of the order of $2 \mu m$) in the problem. The
inverse of this length may be interpreted as a spontaneous curvature $c_0$,
measuring a symmetry breaking between the two
sides of the membrane by incorporating different amount of oil
molecules (in some sense is the analog of chirality in
liquid crystals).
\protect
\section{Theory}
One may models this by regarding a membrane layer  as composed of two different layers
separated by a distance $h$ and with different values of surface tension and surface 
elements one with $\sigma_{+}$ and $dS_{+}$the other $\sigma_{-}$ and $dS_{-}$. 
Therefore one has an energy of the form $ \sigma_{+}\int dS_{+} +\sigma_{-}\int dS_{-} $.
 This energy would include all the elastic contributions (curvature and surface tension 
 of the whole layer) \cite{free}. Expanding now both surface elements $dS_{\pm}$ 
 in powers of $h$, with $dS_{\pm}= dl_{\pm}^{1}dl_{\pm}^{2}=
 dl^{1}dl^{2}\bigl(1+h/2R_1)(1+h/2R_2\bigr)$,
one gets $$ (\sigma_{+}+\sigma_{-}) \int dS +(\sigma_{+}-\sigma_{-})h\int \left( \frac{1}{R_1} +
\frac{1}{R_2}\right) dS .$$ 
Here $\frac{1}{R_1} + \frac{1}{R_2}$ is the mean curvature and $dS=dl^{1}dl^{2}$ is a 
geometrical neutral surface element  between the two layers.
Finally, adding a ``bending'' energy of a tube, the free energy is \cite{granek1}:

\begin{equation}
F =  \frac{\kappa}{2}\int dS \left( \frac{1}{R_1} +
\frac{1}{R_2}-c_0\right)^2 + \sigma\int dS.
\label{energy}
\end{equation}
Where $\kappa$ is the rigidity, $\sigma=(\sigma_{+}+\sigma_{-})-\kappa c_0^2/2$ the
``effective'' surface tension of the bilayer and $c_0=(\sigma_{+}-\sigma_{-})h/4\kappa$ 
is the spontaneous curvature of the layer and is null when the bilayer is symmetric, or when
the neutral surface of the layer is in the middle of the layer
\cite{helf}. Here the spontaneous curvature is not introduced as a parameter, 
but as an assymetry in the layer and a shift of the neutral surface from the middle 
of the layer. Also the surface tension is expressed as combination of the different mechanical
parameters of the mambrane.For an axisymmetric surface characterized by a local
radius $R(z)$ the mean curvature is
\begin{equation}
\frac{1}{R_1} + \frac{1}{R_2} = \frac{1}{ R
\sqrt{1+  R'^2}} - \frac{R''}{(1+  R'^2)^{3/2}}
\label{mean}
\end{equation}
on the other hand the surface element is
$dS = 2\pi  R \sqrt{1+  R'^2 }\,dz.$ Here $R'=d R/d z$

Let $R_0$ be the radius of a non perturbed tube; as in
\cite{chandra} we put
\begin{equation}
R(z) = R_0 \left[\sqrt{1- |u_q|^2}+ \left(\frac{u_q}{\sqrt{2}}e^{i q z}
+cc. \right)\right]
\label{conserv}
\end{equation}
(the square root appears because of conservation of the
volume of the tube, note that $u_{q=0}$ cannot hold a value diffrent from zero)
into the free energy getting  a power  expansion in $u_q$:
\begin{equation}
F = F_0 + \alpha(q)  |u_q|^2 + \frac{\beta(q)}{2} |u_q|^4
\label{free},
\end{equation}
here
$$
F_0 =  \frac{\pi L\kappa}{R_0} ( (1-c_0R_0)^2 + 2  {\cal S}),
$$
${\cal S}\equiv\frac{\sigma}{\kappa c_0^2}$
is a dimensionless number, and the  coefficient of
the quadratic term is
\begin{eqnarray}
\alpha(q) = &\frac{\pi L \kappa}{2R_0}( - (c_0 R_0)^2 -
                (qR_0)^2 - 4(c_0 R_0)(qR_0)^2  \nonumber \\
            & + (c_0 R_0)^2(qR_0)^2 + 2(qR_0)^4  \nonumber \\
            & - 2{\cal S} (c_0R_0)^2(1-(qR_0)^2) ) .
\label{alfa}
\end{eqnarray}

\begin{figure}[tbh]
\centerline{\epsfxsize=80mm \epsfysize=70mm \epsfbox[72 322 540 520]{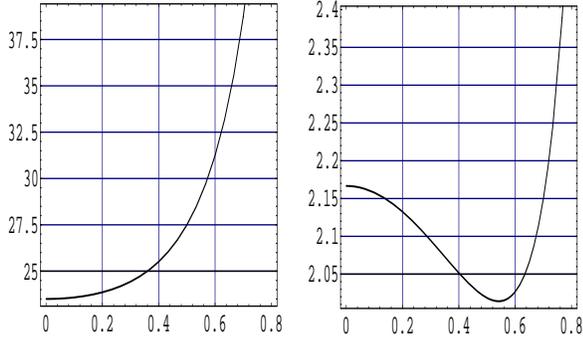}}
\caption{The order parameter $\cal S$ as a function of the product $qR_0$,
the figure in the left is obtained for a zero spontaneous curvature, and the
one in the right is for $c_0 \ne 0$.   }
\label{param}
\end{figure}

The linear instability appears for $\alpha(q)<0$. Geometrically,
the transition $\alpha(q)=0$ allows us to express the control
parameter ${\cal S}$ as a function of the pertubation wavenumber $qR_0$. For
$0<c_0R_0<1/2$,
${\cal S}$ has a minima at $q=0$. The cylinder is stable for low
values of ${\cal S}$, however, as one increases ${\cal
S}$ up to a critical value (${\cal S}(q=0)$), the tube suffers a
long wavelength
instability, without wave number selection. For $1/2<c_0R_0<1$,
${\cal S}$ has a minima for $q=q_c$. A short wavelength instability
appears as soon as ${\cal S}>{\cal S}_c$ see figure (\ref{param}).
From the expresion (\ref{alfa})
one get explicitly the threshold of the instability, hence the wave
vector of the most instable mode ($q_c$), and the value of the
controle parameter (${\cal S}$) at the onset of the instability.

\begin{figure}[tbhp]
\centerline{\epsfxsize=80mm\epsfbox[72 235 540 560]{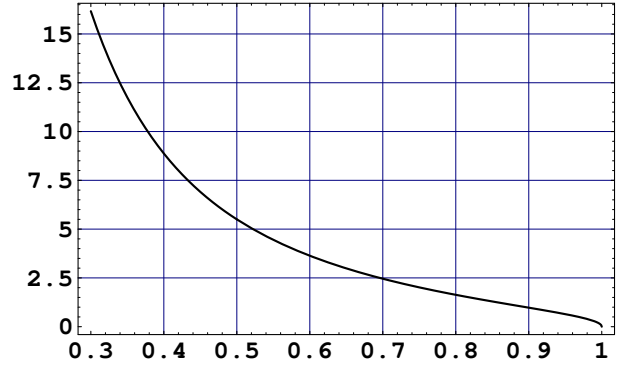}}
\caption{The critical value ${\cal S}_c$ vs. the product $c_0R_0$. As the
radius of the tube approaches spontaneous curvature the tension vanishes as
it should be.}
\label{tensioncritic}
\end{figure}

In the figure (\ref{tensioncritic}), one can see that the control
parameter vanishes as soon as the radius of the tube is equal to the
spontaneous curvature, i.e. when the tubules layer are symmetrical.
From \ref{alfa} and imposing $\alpha(q) = d{\cal S}/ dq=0$ one gets the
critical wavenumber and the threshold. Being the first
\begin{equation}
q_c R_0 = \sqrt{1\pm\sqrt{2}\sqrt{1-c_0R_0}}.\label{qc}
\end{equation}
Note that there are two different solutions, however the sign $+$ represents a
situation where $q_cR_0>1$ and this instability is possible only if
$\cal S$ takes negative values\footnote{Note that this negative surface energy
makes instability even for $c_0=0$.}, however, the selected wavenumber in this
case has a negative ``mean slope'', that is, is a line essentially
perpendicular to the experimental line given in figure \ref{adda1}. The critical
surface tension for the case $q_cR_0<1$ is
\begin{equation}
{\cal S}_c  =  \frac{-3+ 2^{5/2}\sqrt{1-c_0R_0} + 4 c_0R_0 -
(c_0R_0)^2}{2(c_0R_0)^2}.\label{sc}
\end{equation}

The fit represented in figure \ref{adda1} was done using the expression
(\ref{qc}), i.e. we tried $q_cR_0 \sim
\sqrt{1-K\sqrt{1-c_0R_0}}$, the fit gives $K \approx 1.2$ (instead
of 1.4) and $c_0 = 0.16 \mu m^{-1}$  this value is in the range of value measured
for vesicles. In figure (\ref{param}) the value of
${\cal S}_c$ at the $q_c$ is of the order of 2.

At the critical point $q_c$, the sign of the coefficient
$\beta(q_c,{\cal S}_c)$ of the fourth power term $|u_q|^4$, is positive
for all values of $c_0R_0$ such that $1/2<c_0R_0<1$. Therefore
this peristaltic instability is supercritical (a second order transition).
It is interesting to note that for large tubular radius there
is no instability, as we can see in the expression (~\ref{qc}) where $q_c$
becomes a complex number. This characteristic is indeed observed in
the experiment \cite{mathese}. On the other hand,
for small tubules radius, the instability grows for $q_c =0$,
however this is not possible because of the conservation of volume,
in other words a homogeneous growth of the $u_q$ is forbidden.
In conclusion in both cases: i) for small tubular radius:
$R_0<1/c_0$ (experimentally we get $R_0< 1 \mu m$) and for ii) large
tubular radius $R_0>1/c_0$ (experimentally $R_0> 10 \mu m$)
the tubules are stables.
\protect
\section{conclusion}

The oil when incorporated in the membrane of the vesicle, is
distributed assymtrically between the different layers composing the
membrane, i.e. between the outer layer and the inner layer, giving rise to
a non-zero spontaneous curvature. Also a possible scenario would be that
when the swollen micelles reach the tubular vesicles, they load their amphiphiles
at the outer surface and a symetry bearking occures between the outer and the inner layers.
A microscopic description of
the distribution of molecules in terms of the volume fraction of oil
introduced into the alipahtic part of the layer, is necessary to understand
this effect on the spontaneous curvature of the tubules \cite{fourcade2}.  One can 
wonder of how it is possible to vary the spontaneous curvature, without changing the length of the 
surfactant molecule? As for AOT monolayers at the planar interface between the 
oil and brine phases, this can be done by changing the brine salinity \cite{hamid1}, but in the range
of the salinity one can probe, the Debye length  changes from 0.456 nm to
0.608 nm, which is not sufficient to  change the spontaneous curvature of the
tubular vesicles radius  in the range 5 to 20 $\mu m$. This can be perfomed using 
optical tweezers to bring proteins to a particular region of the membrane.

In conclusion, tubules can be excited by different manners: from UV radiations to
laser tweezers; the latter represents a powerfull tool to control the onset of
the instability and the size of the pearls nucleated \cite{moses}. However, it is
found that the  incorporation of another molecule in this case an alkane in the
bilayer,  a more or less good solvent for the aliphatic chain of the surfactant,
the tubules  suffer an instability similar to the raleigh instability.
Therefore, the non-zero spontaneous curvature is created by the
incorporation of oil molecules (dodecane) into the membrane of vesicles,
inducing a short wavelength instability to appear. In our model a surface
tension of the whole layer is necessary but not sufficient
to describe the peristaltic states in fluid tubular membranes. However it is important
to introduce an effective surface tension (which is probably created by the incorporation 
of oil molecules between the aliphatic tails of the surfactant) composed of the true surface
tension of the outer and inner layer which suffer elongation and compression respectively
and the rigidity of the layer as well as its thickness. 

\protect
\section*{acknowledgments}
We are  very grateful to X. Michalet and J. Meunier for help\-ful discussions and
critical comments. Laboratoire de Physique Statistique is assci\'e au CNRS, aux
Universit\'es Paris VI et Paris VII.

\bibliographystyle{unsrt}

\end{document}